\RequirePackage{fix-cm}
\documentclass[twocolumn,epjc3]{svjour3}  
\smartqed  
\RequirePackage{graphicx}
\usepackage{amsmath}
\usepackage{hyperref}  
\usepackage{xcolor}
\journalname{Eur. Phys. J. C}
\begin{document}

\title{\boldmath Neutrino production in the central dark-matter spikes of active galaxies}


\author{Polina Kivokurtseva\thanksref{e1,addr1,addr2}}

\thankstext{e1}{e-mail: kivokurtceva.pi19@physics.msu.ru}

\institute{Faculty of Physics, M.V. Lomonosov Moscow State University,\\ 1-2 Leninskie Gory,  Moscow 119991, Russia \label{addr1}
           \and
           Institute for Nuclear Research of the Russian Academy of Sciences,\\ 60th October Anniversary Prospect 7a, Moscow 117312, Russia \label{addr2}}

\date{Received: date / Accepted: date}

\thankstext{}{INR-TH-2024-019}
\maketitle

\begin{abstract}
Recent multi-messenger observations suggest that high-energy neutrinos may be produced close to central black holes in active galaxies. These regions may host dark-matter (DM) spikes, where the concentration of DM particles is very high. Here we explore the contribution of the DM annihilation to the target photons for the neutrino production, proton-photon interactions, estimate the associated neutrino spectrum and figure out possible future tests of this scenario.
\end{abstract}
\section{Introduction}
\label{sec:intro}
Dark matter (DM) remains one of the most mysterious subject in modern science. Despite the fact that DM makes up the majority of the matter in the Universe, very little is known about it. Numerous direct and indirect detection experiments and collider searches in past years have been aimed to detect dark matter particles, see e.g. Ref.\cite{Cebri_n_2023} for a recent review. Many of them were searching for weakly interacting particles with a mass in the GeV-TeV scale. The lack of success suggests that the mass of DM particles may lay in sub-GeV region. Previously the cosmic bound has disfavored DM fermions with masses below GeV \cite{Bernabei_2008,Undagoitia_2015}. However, a number of new DM models have been established recently \cite{Knapen_2017,Berlin_2018,D_Agnolo_2015,Bertuzzo_2017,Darm__2018}, and planned experiments \cite{Buonocore_2020} provide a strong opportunity to investigate sub-GeV DM. A major obstacle for models of sub-GeV dark matter is that they must meet stringent constrains on the DM annihilation cross section derived from searches for X-ray emission of the Milky Way galaxy \cite{Cirelli_2023} and the Cosmic Microwave Background (CMB) \cite{2020,Slatyer_2009}.

In the center of a bright active galactic nuclei (AGN), there is a supermassive black hole (SMBH) with a typical mass of order $10^8$ solar masses. It was established through numerical calculations that if dark matter is present at the galactic center it will be redistributed into 'spike'\cite{Gondolo_1999,PhysRevLett.92.201304,PhysRevLett.93.061302}. The concentration of dark-matter particles in the spike is huge, which opens more possibilities for indirect detection.

A lot of progress have been made in detection of high-energy neutrinos, with the existence of their astrophysical flux established by IceCube, ANTARES and Baikal-GVD experiments, see e.g. Ref. \cite{Troitsky:2023nli} for a recent review. While it is non-trivial to determine their precise astrophysical origins, it has long been known that high-energy neutrino sources such as active galactic nuclei (AGN) hold tremendous potential. Various theories of neutrino production have been proposed recently, however the precise mechanism is still unknown. Depending on the source, the neutrino production mechanism can vary. 

Recent data from IceCude has shown neutrino emission from NGC 1068 in the energy range 1-20 TeV. Evidence of high-energy neutrinos and gamma rays from NGC 1068 \cite{2022} points the site of neutrino production to the supermassive black hole proximity. This suggests that the dark-matter spike may affect neutrino production.

In the present work, we aim to discuss a mechanism through which dark matter can be connected with the neutrino production, keeping NGC 1068 as a working example. Electrons from DM annihilation can upscatter ambient photons, which will act as target photons in $p\gamma$-processes, which produce high-energy neutrinos.

The rest of the paper is organized as follows. In Sec.~\ref{sec:mod}, we give an overview of the dark matter spikes near SMBH and of the mechanism of dark matter interaction with standard model (SM). We start in Sec.~\ref{sec:mod:spike} with a brief review of the dark matter distribution around SMBH. We then concentrate on various models of DM interaction with standard model in Sec.~\ref{sec:mod:inter}. Section Sec.~\ref{sec:flux} contains  detailed numerical calculations of the fluxes. In Sec.~\ref{sec:diss},
we compare the mechanism we propose with previous studies and discuss its observational implications. We briefly conclude in Sec.~\ref{sec:concl}.
\section{Overview and estimates}
\label{sec:mod}
\subsection{Spike phenomenology}
\label{sec:mod:spike}
 The mechanism of black hole formation is still poorly understood. One of scenarios is adiabatic growths from a small seed all the way to SMBH, which reside in AGN center. In such model spike of dark matter may be formed \cite{Gondolo_1999,Sigurdsson_1995,Ullio_2001}. If dark matter halo with singular power-law profile $\rho(r) = \rho_0(r/r_0)^{-\gamma}$, with $0<\gamma<2$, $\rho_0$ a reference
density at $r = r_0$, is present in the center, than it will be evolved into:
\begin{equation}
    \rho_{\textup{sp}}=\rho_R \cdot g_{\gamma}(r)\left(\frac{R_{\textup{sp}}}{r}\right)^{\gamma_{s p}}
\end{equation}
 because of black hole growth. The size of the spike $R_{\textup{sp}}=\alpha_{\gamma}r_0(M_{\textup{BH}}/\rho_0 r_0^{3})^{1/3-\gamma}$, where $M_{\textup{BH}}$ is mass of black hole, can be calculated through obtaining the normalization $\alpha_{\gamma}$ and the factor $g_{\gamma}(r)$ numerically. The slope of the spike is given by $\gamma_{\textup{sp}}=(9-2\gamma)/(4-\gamma)$. For $0<\gamma<2$, the function $g_{\gamma}(r)$ can be calculated as $g_{\gamma}(r) \approx (1-4R_{s}/r)$ , with $R_S$ the Schwarzschild radius. Normalization factor $\rho_R=\rho_0(R_{\textup{sp}}/r_0)^{-\gamma}$ was selected for consistency with the density profile outside the spike. The density of the spike falls rapidly to zero at $r < 10R_{\textup{S}}$, vanishing for $r < 4R_{S}$. Although a more precise analysis of the adiabatic growth of the dark matter spike, taking relativistic factors into account, reveals that the spike in fact disappears at $r = 2R_S$ rather than $4R_S$ and that the density of dark matter particles increases significantly close to the core \cite{PhysRevD.88.063522}. For our calculations we assume the NFW profile for the initial DM distribution with $\gamma=1$.

 Making assumption that particles of dark matter can annihilate, the maximal dark matter density in the inner regions of the spike is saturated to $\rho_{\textup{sat}} = m_{\textup{DM}}/(\langle\sigma v\rangle t_{\textup{BH}})$, where $\langle\sigma v\rangle$ is the velocity averaged dark matter annihilation cross section, and $t_{\textup{BH}}$ is the age of SMBH and $m_{\textup{DM}}$ is dark-matter particle mass. The full DM density profile can be written as:
\begin{equation}
    \rho_{\textup{DM}}(r) = \begin{cases}
  0,  & r<4R_{\textup{S}} \\
  \frac{\rho_{\textup{sp}}\rho_{\textup{sat}}}{\rho_{\textup{sp}}+\rho_{\textup{sat}}}, & 4R_{\textup{S}}<r<R_{\textup{sp}} \\
  \rho_{0}(\frac{r}{r_{0}})^{-\gamma}(1+\frac{r}{r_{0}})^{\gamma - 3} , & r> R_{\textup{sp}}
\end{cases}
\end{equation}

\begin{figure}[h]
	\centering
	\scalebox{0.5}{\includegraphics{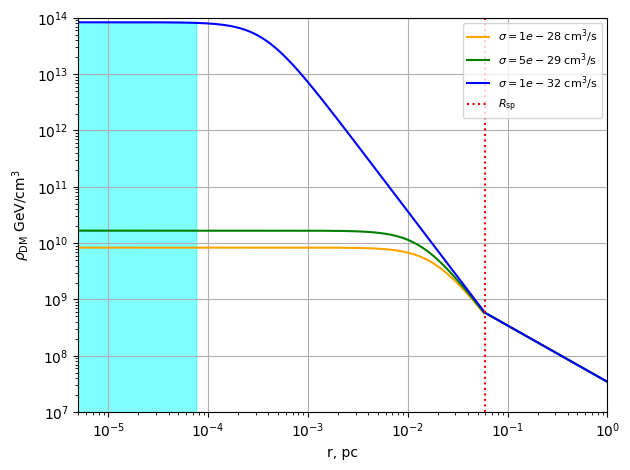}}
    \caption{Dark matter distribution around the SMBH of NGC 1068 for different values of the DM
self-annihilation cross section, the shaded area correspond to the area of neutrino production.}
	\label{dencity} 
\end{figure}

We determined the normalization $\rho_0$ for this model through the uncertainty on the black hole mass \cite{PhysRevD.82.083514,PhysRevD.96.063008}. For our calculations we take $t_{\textup{BH}}=10^{10}$ \cite{10.1093/mnras/staa3363} and $r_0= 10$ $\textup{kpc}$. Numerically for NGC 1068 we get $\rho_0=1.23$ $\textup{GeV}/\textup{cm}^{3}$.The DM profiles for NGC 1068 are
shown in the Fig.\ref{dencity}. Neutrino production appears to occur in an area smaller than $(30 - 100)R_{\textup{S}}$, according to multi-messenger data \cite{Murase_2022}. Density of DM is high in this region. This makes it interesting to discuss a possible connection between dark matter and neutrino generation.

\subsection{Dark matter interaction with SM}
\label{sec:mod:inter}
The two primary models of fermionic dark matter are Majorana DM \cite{Ikemoto_2023} and Dirac fermion DM\cite{PhysRevD.88.036011}. The primary distinction between these two models is how DM scatters.
 The standard model (SM) is acknowledged to have no room for dark matter \cite{HOOFT1971167}. Nonetheless, over the previous few decades, a number of options to expand SM have come forth. To set the stage for our computations, we  briefly mention a number of dark matter models in this section. 

The SM, the DM, and the fields mediating the DM's interactions with the SM are the three sectors into which any dark matter model can be split. The idea that the DM communicates through some sort of "portal" with the visible, Standard Model (SM) sector is highly motivated \cite{alexander2016darksectors2016workshop}. Still up for debate, though, is what constitutes a mediating state. There are several potential "portals" available: Higgs portal, see e.g. Ref. \cite{Arcadi_2020} ,kinetic mixing portal, see e.g. Ref. \cite{Chun_2011}, fermion portal, see e.g. Ref.\cite{Bai_2013}, etc.. 

\begin{figure}[h]
	\centering
	\scalebox{0.4}{\includegraphics{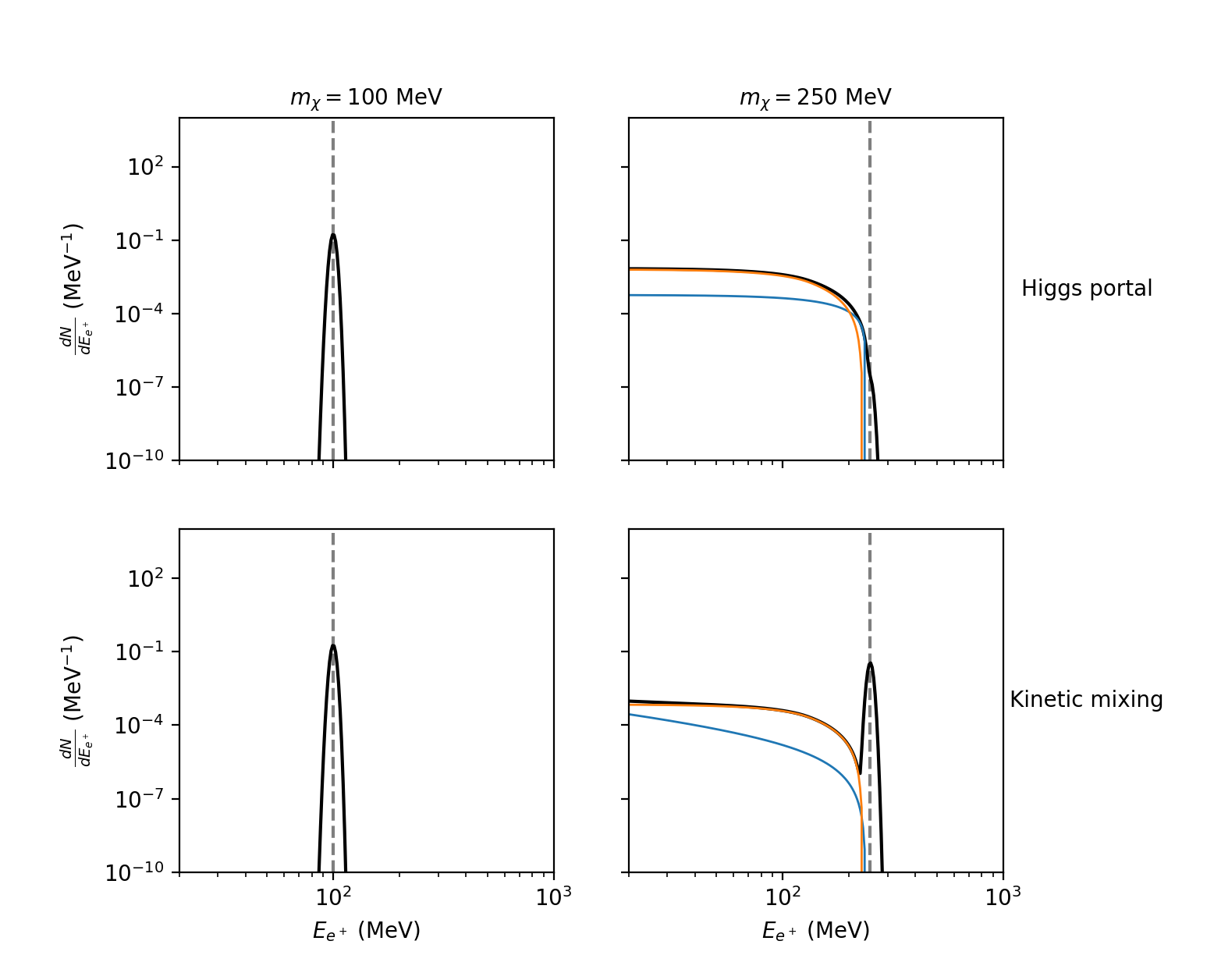}}
    \caption{Positron annihilation spectra from the scalar model with Higgs portal couplings and vector model with kinetic mixing couplings. The grey vertical dashed line indicates the location of the monochromatic dark matter annihilation. Colorful lines correspond to $\pi^{+}\pi^{-}$ and $\mu^{+}\mu^{-}$ decay .}
	\label{KM_HP} 
\end{figure}

Dark-matter particles with sub-GeV masses can annihilate \cite{Boehm_2004}. Only a few annihilation channels are permitted due to the DM mass range \cite{PhysRevD.92.023533}. We will mostly pay attention to one main channels of annihilation:
\begin{equation}
    \chi \chi \rightarrow e^{+} e^{-} 
\end{equation}
which is kinematically open whenever $m_{\textup{DM}} > m_{\textup{e}}$. For the following calculations we will use open-source code "Hazma" \cite{Coogan_2020}. The code allows one to simulate the annihilation of sub-GeV dark matter into different final states. It includes all relevant particle interactions for various kinematic regimes. Furthermore, the final particle spectrum consists of a continuum piece from unstable particle decays as well as a line-like spectrum from dark matter annihilation. Specifically, this enables a departure from the $\delta$-functional approximation. Hazma employs various models to illustrate how DM and SM can interact. Calculated electron spectra is shown in Fig.\ref{KM_HP}.

To obtain the electron spectra from DM annihilation, we lay constrains on model parameters. They were chosen so that annihilation cross section will be aligned with present constraints, which were derived by using low energy measurements by Voyager 1 \cite{Boudaud_2017} and constrains from the CMB \cite{Slatyer_2016}.
For all of the following calculations we take  $m_{\textup{DM}}=250$ MeV and dark matter self-annihilation cross section is $\langle\sigma v\rangle = 5\cdot 10^{-29} \textup{cm}^{3}/\textup{s}$.
\section{Neutrino flux}
\label{sec:flux}
The most general mechanism of high-energy neutrino production involves $p\gamma$ or $pp$ processes. However, the location where these reactions take place, and details of the mechanism are still up to debate. NGC 1068, among others, is a promising source of PeV neutrinos \cite{PhysRevD.94.103006}. To get neutrinos of such energies, protons should scatter on target photons with keV to MeV energies. In this paper we propose another mechanism: energetic electrons and positrons from DM annihilation could upscatter ambient photons to needed energies.

\subsection{Diffusion-loss equation }
\label{sec:flux:diff}
After DM annihilation, electrons and positrons interact with surrounding environment and lose their energy through Coulomb interactions and ionization($b_{\textup{ion+Col}}$), bremsstrahlung($b_{\textup{brem}}$), synchrotron radiation($b_{\textup{syn}}$) and Inverse Compton scattering($b_{\textup{ICS}}$). So, we can write the energy-loss function as follows:
\begin{equation}
    -\frac{dE}{dt}=b_{\textup{ICS}}+b_{\textup{syn}}+b_{\textup{brem}}+b_{\textup{ion+Col}}
\end{equation}
 Let us briefly comment on each term.

 Losses by Coulomb interactions and ionization as well as bremsstrahlung occur on both neutral and ionized matter. They depend on the gas density in the surrounding area which is usually determined from astrophysical observations. 
For $b_{\textup{brem}}+b_{\textup{ion+Col}}$ we follow the choose of parameters for gas density in \cite{Eichmann_2022}.

Losses by Inverse Compton Scattering occur on Inter-Stellar Radiation Field (ISRF) components: CMB, optical star-light and dust-diffused infrared light. Exact descriptions for the optical and infrared components need accurate maps. Exact form of ICS losses reads \cite{Buch_2015}:
 \begin{equation}
    \begin{split}
& b_{\rm \textup{ICS}} = \\
& 3c\, \sigma_{\rm T} \int_0^\infty d\epsilon\, \epsilon \int_{1/4\gamma^2}^1dq\ n(\epsilon) \frac{(4\gamma^2-\Gamma_\epsilon)q-1}{(1+\Gamma_\epsilon q)^3}[ 2q\ln q + \\
& +q+1-2q^2+\frac{1}{2}\frac{(\Gamma_\epsilon q)^2}{1+\Gamma_\epsilon q}(1-q) ], 
\end{split}
\label{eq:enlossICS}
\end{equation}
where $n(\epsilon)$ is the number density  of photons of the ISRF, with energy $\epsilon$, $\gamma = E/m_\textup{e}$ is the relativistic factor of the electrons and positrons and $\Gamma_\epsilon = 4\epsilon\gamma/m_\textup{e}$.
In the Thomson limit losses reduce to : 
\begin{equation}
b_{\rm \textup{ICS}} = \frac{4\,c\ \sigma_{\rm T}}{3\, m_e^2}\, E^2  \int_0^\infty d\epsilon\ \epsilon\ n(\epsilon) ,
\label{eq:enlossICSThomson}
\end{equation}

Synchrotron losses are present, when charged particles move through a magnetic field. They can be calculated as follows \cite{Buch_2015}:
\begin{equation}
   b_{\rm \textup{syn} } = \frac{4\,  c \ \sigma_{\rm T}}{3\, m_e^2}\, E^2\, \frac{B^2}{8\pi} ,
\label{eq:enlosssyn}
\end{equation}
where $B$ is the strength of the magnetic field. In our calculations, we adopt a fiducial magnetic field value of $B=0,1$ G. This choice is motivated by studies showing a wide range of field strengths in active galactic nuclei. For instance, \cite{Fang_2023} estimate values up to 0.5 G in compact jet regions, while \cite{inoue2024highenergyneutrinosgammarays} use lower values in more central areas such as the torus. Our focus is on the dark matter spike near the black hole, where we expect the field to lie between these two extremes. The adopted value thus represents a reasonable and physically motivated average.We take radiation density of CMB as a blackbody spectra with $T=2.753 \textup{K} $. 
To obtain the number density of electrons we need to solve the standard diffusion-loss differential equation
\begin{equation}
   \frac{\partial f}{\partial t} - \nabla (K(E, r)\nabla f) - \frac{\partial }{\partial E}(b(E,r)f)  = Q(E,r) ,
\end{equation}
where the electron source term  can be written as:
\begin{equation}
    Q_{\textup{e}}=\frac{1}{2}(\frac{\rho}{m_{\textup{DM}}})^2\langle \sigma v \rangle \frac{dN_{e}}{dE} , \label{eq:1}
\end{equation}
The $dN_{e}/dE$ is electron spectrum which is calculated in Sec. \ref{sec:mod:inter}, $\rho$ is dark matter distribution(spike) and $\langle \sigma v \rangle$ annihilation cross section.
\newline
Let us neglect diffusion and use 'on the spot' approximation \cite{Cirelli_2009}. Then we can calculate the electron density as: 
\begin{equation}
    \frac{dn_{\pm}}{dE_{\textup{e}}}(E_{\textup{e}},\Vec{x})=\frac{1}{b_{tot}(E_{\textup{e}},r)}\int_{E_{\textup{e}}}^{m_{\textup{DM}}}d\widetilde{E}Q_{\textup{e}}(\widetilde{E},r) .
\end{equation}

\subsection{Electron-photon interaction}
\label{sec:flux:elec}
As previously mentioned, electrons must up-scatter photons existing in the neutrino production area in order to produce neutrinos (see Fig.~\ref{dencity}) with the observed energy. We must first obtain the target photons that protons will interact with. In order to carry out a thorough computation of the neutrino flux for the selected galaxy. 

To do so, we use the TransportCR code \cite{kalashev2014simulationsultrahighenergy} based on transport equations. The code allows one to simulate the propagation of nuclei and nucleons through the media filled with arbitrary photon backgrounds, also tracing secondary electron-photon cascades and neutrinos from their interactions. The background spectrum can be supplied in the table form. The code includes all relevant particle interactions for various kinematical regimes.

\begin{figure}[h]
	\centering
	\scalebox{0.7}{\includegraphics{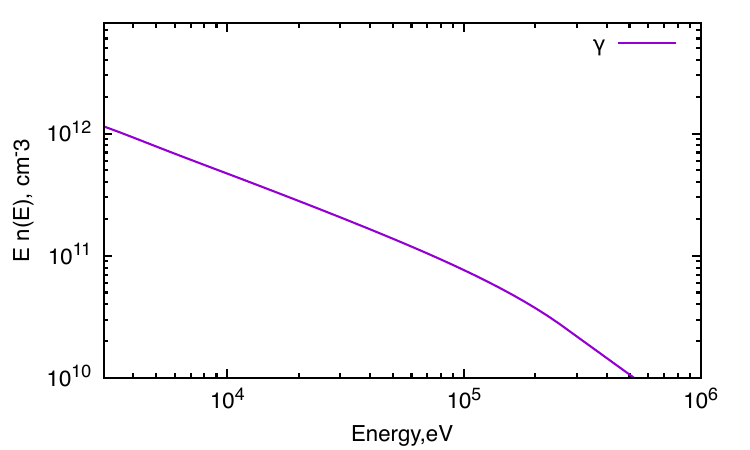}}
    \caption{Target photons resulting from the interaction of electrons from dark matter with the ultraviolet component of the spectral energy distribution}
	\label{targetUV} 
\end{figure}

We take calculated in sec.~\ref{sec:flux:diff} electron density from dark matter annihilation and the optical and ultraviolet (UV) components of the spectral energy distribution (SED) emitted by the accretion disks of Seyfert galaxies \cite{1999} as the target photon fields for electrons. The optical/UV photons are targets for inverse Compton scattering, where relativistic electrons transfer energy to the photons. The results of these simulations are presented in Fig. \ref{targetUV}.

Active galactic nuclei also provide MeV photons via SM processes (non-thermal coronal Comptonization, CR interactions). 
We parametrize the MeV tail of the corona by the fraction $f_{\rm MeV}\equiv L_{\rm MeV}/L_{\rm cor}$. 
The corresponding MeV photon number density at radius $R$ is
\begin{equation*}
    n_{\gamma}^{\rm SM}(R)\;\simeq\;\frac{f_{\rm MeV}\,L_{\rm cor}}{4\pi R^2\,c\,\langle\varepsilon\rangle},
    \qquad \langle\varepsilon\rangle\simeq 1~{\rm MeV}.
\end{equation*}
Adopting $f_{\rm MeV}=10^{-2}$ and $L_{\rm cor}=1\times10^{43}\,{\rm erg\,s^{-1}}$ (e.g., non-thermal hybrid corona arguments \cite{Inoue_2007}), 
 we find
\[
n_{\gamma}^{\rm SM}\approx 7.61\times10^{6}\ \mathrm{cm^{-3}},  R=50R_S,
\]
From Fig.~\ref{targetUV} we read the DM-induced photon density near 1 MeV for our benchmark 
$\langle\sigma v\rangle_0=5\times10^{-29}\,{\rm cm^3\,s^{-1}}$ as $n_{\gamma}^{\rm DM}\approx 10^{10}\ \mathrm{cm^{-3}}$.
Away from deep saturation, $n_{\gamma}^{\rm DM}\propto\langle\sigma v\rangle$, so the cross section required for equality with the SM field is
\begin{equation*}
    \langle \sigma v \rangle_{\rm req} \;=\; \langle \sigma v \rangle_0\;
    \frac{n_{\gamma}^{\rm SM}}{n_{\gamma}^{\rm DM}}\, .
\end{equation*}
Numerically, this gives 
$\langle\sigma v\rangle_{\rm req}\approx1.5\times10^{-32}\,{\rm cm^3\,s^{-1}}$ .

\subsection{Neutrino flux}
\label{sec:flux:obj}
 For the target photon spectrum, we use spectra from previous section \ref{sec:flux:elec}. To simulate interactions we use already mentioned TransportCR code \cite{kalashev2014simulationsultrahighenergy}. 

For our calculations we use parameters of a well known active galaxy NGC 1068. This source was associated with a
high-energy neutrino event detected by IceCube \cite{2022}. 

In this work we do not specify the precise location of the zone where the protons are accelerated, nor a particular acceleration mechanism.The required proton energies can be obtained by acceleration through magnetic reconnection \cite{10.1093/mnras/stv2337},  by acceleration in the electromagnetic field close to the black hole \cite{Ptitsyna_2016}.

The simulation is performed in the frame of the source. We assume the power-law injection spectrum of protons with energies $E_{\textup{p}}$ up to $10^{15}$ eV with sharp cutoff and
spectral index $\alpha=2.7$. The fact that the outcomes are insensitive to plausible parameter adjustments justifies the selection of the highest energy. The magnetic field is assumed to be 0.022 G, as reported in \cite{Fang_2023}.

Since the proton luminosity cannot be directly inferred from observations, we treat it as a free parameter in our model. For normalization, we consider two scenarios: one where the proton luminosity is scaled to the Eddington luminosity ($L_{\textup{Edd}}$) and another where it is set to $100L_{\textup{Edd}}$. This range allows us to explore the potential impact of varying proton luminosities on the resulting neutrino flux.

\begin{figure}[h]
	\centering
	\scalebox{0.7}{\includegraphics{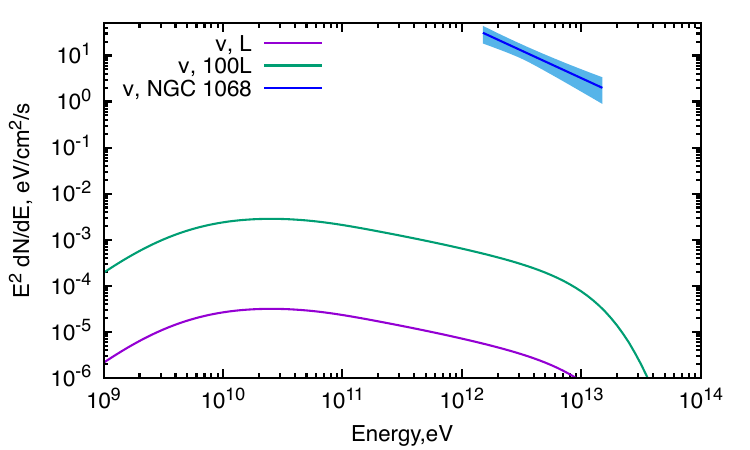}}
    \caption{Predictions of the observed spectra of neutrino ($\nu_{\mu}+\bar\nu_{\mu}$) from NGC 1068 (the disc-related component). The green line represents proton luminosity normalized by $L_{\textup{Edd}}$, the pink line represents proton luminosity normalized by $100L_{\textup{Edd}}$ and  dark blue line shows the best-fit neutrino spectrum, and the corresponding blue band covers all powerlaw neutrino fluxes that are consistent with the data at 95$\%$ C.L. from IceCube data.}
	\label{neutrinoUV} 
\end{figure}

The resulting spectrum of neutrinos produced at the source is shown in Fig. \ref{neutrinoUV}. Our calculations indicate that the predicted neutrino flux is several orders of magnitude lower than the flux detected by IceCube \cite{NGC1068_2022}. We show neutrino flux from dark matter annihilation for individual source to stress out that this mechanism can contribute to overall neutrino flux from individual source but can not explain the whole flux.

Our goal in this paper was to show neutrino flux which originate from dark matter spike around black hole, so we do not take into account other photon sources (see for example \cite{Das_2024}) which can contribute to neutrino flux.

In our setup the neutrino flux scales approximately as $\Phi\sim\langle\sigma v\rangle L_{\textup{p}}$. Using the IceCube measurement for NGC 1068 \cite{NGC1068_2022} and the deficit seen in Fig.~\ref{neutrinoUV}, we translate the shortfall into a “required’’ cross section: $\langle\sigma v\rangle_{\textup{req}}=F\langle\sigma v\rangle_{0}$, where $F$ is the ratio of the IceCube band to our prediction. From the figure we read $F\approx (3.5-3.8)\cdot 10^{-6}$ for the purple line. This implies $\langle\sigma v\rangle_{\textup{req}}=(1.7-1.9)\cdot10^{-22} \textup{cm}^3/\textup{s}$. For $m_{\textup{DM}}=250$ MeV Planck CMB data imply $\langle\sigma v\rangle \leq (1.3-2.0)\cdot 10^{-28} \textup{cm}^3/\textup{s}$ \cite{Slatyer_2016}. The comparison remains valuable, however, because the CMB and spike environments probe energy injection at very different redshifts and characteristic velocities, and are therefore complementary. 

\section{Discussion}
\label{sec:diss}
Multi-messenger data point to the generation of high-energy particles close to SMBH. A portion of the neutrino flux may be connected to the dark matter, which is expected to be densely concentrated in these regions. In this study, we have meticulously investigated this potential connection.

Numerous Seyfert galaxies within the BASS (BAT AGN Spectroscopic Survey) collection exhibit characteristics that deviate significantly from the model parameters designed to explain the observed neutrino emission from NGC 1068 \cite{abbasi2024icecubesearchneutrinoemission}. This misalignment suggests that the mechanisms driving neutrino production in NGC 1068 may not be universally applicable to all Seyfert galaxies, or that additional physical processes are at play. For instance, variations in the accretion dynamics, jet properties, or dark matter distributions around the central SMBHs could contribute to these discrepancies. These findings highlight the complexity of high-energy particle production in AGN and underscore the need for continued and refined searches for neutrino emission from Seyfert galaxies. While a central dark-matter “spike” is theoretically well motivated in the case of adiabatically growing black holes, its long-term survival is uncertain. Dynamical processes such as stellar scattering and heating can significantly deplete the inner density profile over Gyr timescales, reducing the predicted annihilation signal by orders of magnitude \cite{Merritt_2004}. Observationally, the presence of a canonical spike profile is disfavored both in active galactic nuclei \cite{sharma2025novelmethodtracedark} and in the Milky Way’s Galactic Center \cite{Lacroix_2018}. For this reason, the adoption of an idealized adiabatic spike should be regarded as a benchmark calculation, rather than a robust prediction. 

In this work, we have demonstrated that a significant portion of the neutrino flux from individual sources could originate from dark matter interactions in the vicinity of SMBHs. Our analysis indicates that while this mechanism provides a plausible explanation for a fraction of the observed neutrino flux, it is insufficient to account for the entirety of the detected neutrino emission from NGC 1068. This suggests that additional processes, such as hadronic interactions in AGN jets, coronal emission, or other high-energy astrophysical phenomena, likely play a significant role in neutrino production within Seyfert galaxies.

Nevertheless, the proposed dark matter-driven mechanism is not limited to Seyfert galaxies alone; it has the potential to operate in any galaxy hosting a central black hole. Given the vast number of such galaxies in the universe, this mechanism could collectively contribute to the diffuse neutrino background. While the neutrino flux from individual regular galaxies may be modest, the cumulative effect of many such sources could be substantial. This highlights the importance of considering a wide range of galactic environments when modeling the diffuse neutrino background, even if other mechanisms dominate the neutrino flux in Seyfert galaxies specifically.

\begin{figure}[h]
	\centering
	\scalebox{0.7}{\includegraphics{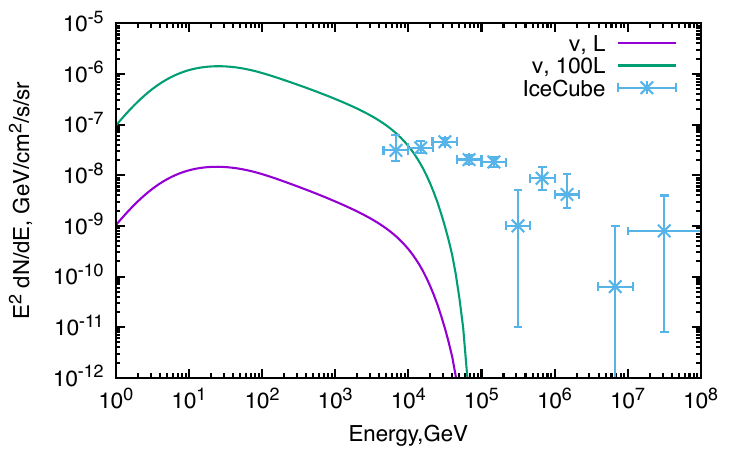}}
    \caption{Predictions of the observed spectra of neutrino ($\nu_{\mu}+\bar\nu_{\mu}$) from AGN source populations. The pink line represents proton luminosity normalized by $L_{\textup{Edd}}$,the green line represents proton luminosity normalized by $100L_{\textup{Edd}}$ and blue points correspond to measurement of the astrophysical diffuse neutrino flux by IceCube}
	\label{dif} 
\end{figure}

To get estimates we calculate the diffuse neutrino contribution from SMBHs. We adopt the SMBH mass and redshift distributions from Ref.~\cite{Pesce_2021} and integrate over redshift 
z to determine the number density of SMBHs at different cosmological distances. For our calculations, we focus on SMBHs with masses $M\approx10^{8}M_{sun}$, which are comparable to the mass of the black hole in NGC 1068. Although we include the redshift range z from 0 to 6, the contribution is dominated by sources at moderate redshifts (
$z<3$) due to the evolution function of active galaxies, which significantly suppresses their number density at higher redshifts. This also justifies our use of a fixed $t_{BH}$ value across all redshifts, as variations at high z do not significantly impact the final result.
The resulting diffuse neutrino flux, along with the observed flux estimates from IceCube \cite{naab2023measurementastrophysicaldiffuseneutrino}, is shown in Fig.~\ref{dif}.By using the lowest point of the IceCube data, we can estimate the proton luminosity to be approximately $50 L_{\textup{Edd}}$ .
The proposed new mechanism of neutrino production has the potential to contribute significantly to the overall neutrino flux. If AGN are capable of producing a substantial population of high-energy protons, the resulting neutrino flux could be considerable.

Figure~\ref{dif} suggests that measurements of the astrophysical diffuse neutrino flux at energies $\sim (1-10)$~TeV are most constraining for the present scenario. Observation of these neutrinos, presently unavailable because of strong atmospheric background, would require high statistics and precise energy reconstruction. Experiments like Baikal-GVD \cite{refId0} and KM3NeT \cite{Adrian-Martinez_2016}, which are currently under construction but already operating, will supplement IceCube in solving this problem. Upper limits on, or measurement of, the astrophysical neutrino flux at these energies would make it possible to constrain this scenario and, consequently, the properties of dark matter. 

\section{Conclusion}
\label{sec:concl}
High-energy astrophysical neutrino production may be linked with dark matter, which can form a dense spike near SMBH. To produce high-energy neutrinos, protons should scatter on target-photons of the keV to MeV energies, which can be boosted to this energies by electrons from dark-matter annihilation. This work estimates the contribution of dark-matter related process to the overall neutrino flux. This mechanism can be tested with observations of astrophysical neutrinos at TeV energies.

\begin{acknowledgements}
The author is very grateful to Sergey Troitsky, Mikhail Kuznetsov, Maria Kudenko for useful discussions. This work is supported in the framework of the State project ``Science'' by the Ministry of Science and Higher Education of the Russian Federation under the contract 075-15-2024-541. 
\end{acknowledgements}
\bibliographystyle{spphys} 

\bibliography{darkM}  

\end{document}